\documentclass[12pt]{article}
\pdfoutput=1
\usepackage{latexsym, graphicx,cite}
\usepackage{amsmath}
\usepackage{amssymb}
\usepackage{amsthm}

\usepackage[top = 1. in,bottom = 1. in, left = 1 in, right=1 in]{geometry}

 \newcommand{\arXiv}[1]{\href{http://www.arXiv.org/abs/#1}{arXiv:#1}}
\usepackage[colorlinks=true, linkcolor=blue, bookmarks=true]{hyperref}

\makeatletter
\renewcommand\section{\@startsection {section}{1}{\z@}%
                  {-3.5ex \@plus -1ex \@minus -.2ex}
                  {2.3ex \@plus.2ex}%
                  {\normalfont\large\bfseries}}
\renewcommand\subsection{\@startsection{subsection}{2}{\z@}%
                   {-3.25ex\@plus -1ex \@minus -.2ex}%
                   {1.5ex \@plus .2ex}%
                   {\normalfont\bfseries}}
\makeatother

\newcommand{\Iden}{\mathcal{D}}

\newcommand{\beq}{\begin{equation}}
\newcommand{\eeq}{\end{equation}}
\newcommand{\ber}{\begin{array}}
\newcommand{\eer}{\end{array}}
\newcommand{\del}{\partial}

\newcommand{\ssty}{\scriptstyle}
\newcommand{\dsty}{\displaystyle}

\newcommand{\de}{\delta}

\newcommand{\ena}{\end{eqnarray}}
\newcommand{\beqa}{\begin{eqnarray}}
\newcommand{\eeqa}{\end{eqnarray}}
\newcommand{\bea}{\begin{eqnarray}}
\newcommand{\eea}{\end{eqnarray}}

\theoremstyle{remark}

\renewcommand{\Re}{\operatorname{Re}}
\renewcommand{\Im}{\operatorname{Im}}

\begin{document}
\begin{titlepage}
\begin{flushright}
\phantom{arXiv:yymm.nnnn}
\end{flushright}
\vspace{1cm}
\begin{center}
{\LARGE\bf Solvable cubic resonant systems}\\
\vskip 15mm
{\large Anxo Biasi,$^{a}$ Piotr Bizo\'n$^{\,b}$ and Oleg Evnin$^{c,d}$}
\vskip 7mm
{\em $^a$ Departamento de F\'\i sica de Part\'\i culas, Universidade de Santiago de Compostela
	 and Instituto Galego de F\'\i sica de Altas Enerx\'\i as (IGFAE), Santiago de Compostela, Spain}
\vskip 3mm
{\em $^b$  Institute of Physics, Jagiellonian University, Krak\'ow, Poland}
\vskip 3mm
{\em $^c$ Department of Physics, Faculty of Science, Chulalongkorn University,
Bangkok, Thailand}
\vskip 3mm
{\em $^d$ Theoretische Natuurkunde, Vrije Universiteit Brussel and\\
The International Solvay Institutes, Brussels, Belgium}
\vskip 7mm
{\small\noindent {\tt anxo.biasi@gmail.com, bizon@th.if.uj.edu.pl, oleg.evnin@gmail.com}}
\vskip 10mm
\end{center}
\vspace{1cm}
\begin{center}
{\bf ABSTRACT}\vspace{3mm}
\end{center}
Weakly nonlinear analysis of resonant PDEs in recent literature has generated a number of resonant systems for slow evolution of the normal mode amplitudes that possess remarkable properties. Despite being infinite-dimensional Hamiltonian systems with cubic nonlinearities in the equations of motion, these resonant systems admit special analytic solutions, which furthermore display periodic perfect energy returns to the initial configurations. Here, we construct a very large class of resonant systems that shares these properties that have so far been seen in specific examples emerging from a few standard equations of mathematical physics (the Gross-Pitaevskii equation, nonlinear wave equations in Anti-de Sitter spacetime). Our analysis provides an additional conserved quantity for all of these systems, which has been previously known for the resonant system of the two-dimensional Gross-Pitaevskii equation, but not for any other cases.

\vfill

\end{titlepage}


\section{Introduction}

We shall be concerned with dynamical systems whose equations of motion are of the form
\begin{equation}
i\dot{\alpha}_n=\hspace{-5mm}\sum_{\begin{array}{c}\vspace{-6mm}\\ \ssty m,k,l=0\vspace{-2mm}\\ \ssty n+m=k+l\end{array}}^{\infty}\hspace{-5mm} C_{nmkl}\bar{\alpha}_m{\alpha}_k\alpha_{l} =\sum_{m=0}^{\infty}\sum_{k=0}^{n+m} C_{nmk,n+m-k}\bar{\alpha}_m{\alpha}_k\alpha_{n+m-k},
\label{eq:flow_eq_linear_spectrum}\vspace{-2mm}
\end{equation} 
with $\alpha_n$ being complex dynamical variables ($n$ ranges from 0 to $\infty$), and dot denoting the time derivative. $C_{nmkl}$ are real number that shall be referred to as the {\em interaction coefficients}. They are symmetric under permutations $n\leftrightarrow m$, $k\leftrightarrow l$ and $(n,m)\leftrightarrow (k,l)$.  We shall assume $C_{0000}\ne 0$, which allows one to rescale the time variable and set
\beq
C_{0000}=1,
\label{C0000}
\eeq
a convention we shall adopt from now on.

Such systems often arise \cite{FPU,CEV1,CEV2,GHT,BMR,GT,CF,BMP,BBCE,BHP,BEL,GGT,BBCE2,BHP2,AdS4} in weakly nonlinear analysis  of PDEs with cubic nonlinearity whose spectrum of frequencies of linearized perturbations is perfectly resonant (the difference of any two frequencies is integer in appropriate units). In fact, the resonance condition $n+m=k+l$ in the sum above precisely reflects the principal role of resonant combinations of frequencies in weakly nonlinear regimes. Examples of such equations are the Gross-Pitaevskii equation describing Bose-Einstein condensates in harmonic traps \cite{GHT,GT,BMP,BBCE,GGT,BBCE2} and various nonlinear problems in Anti-de Sitter spacetime \cite{FPU,CEV1,CEV2,BMR,CF,BEL,AdS4}, in particular, those studied in relation to its conjectured nonlinear instability \cite{BR,rev2}.  In such situations, because of the presence of resonances, amplitudes and phases of linearized modes aquire slow drifts due to effects of nonlinearities (no matter how small the nonlinearities are). The leading part of this drift effect is accurately described by the time-averaging method, which precisely produces an equation of the form (\ref{eq:flow_eq_linear_spectrum}) for the slow evolution of the complex amplitudes of the linearized modes. In this context,  (\ref{eq:flow_eq_linear_spectrum}) is called the `resonant' or the `effective' system. Here, we shall directly focus on the dynamical properties of (\ref{eq:flow_eq_linear_spectrum}) without giving extensive details of how it emerges in weakly nonlinear analysis of PDEs. Interested readers may consult \cite{murdock,KM} for the underlying theory.

The resonant system (\ref{eq:flow_eq_linear_spectrum}) is Hamiltonian with the Hamiltonian
\beq
H=\hspace{-5mm}\sum_{\begin{array}{c}\vspace{-6.5mm}\\ \ssty n,m,k,l=0\vspace{-2mm}\\ \ssty n+m=k+l\end{array}}^{\infty}\hspace{-5mm} C_{nmkl}\bar{\alpha}_n\bar{\alpha}_m{\alpha}_k\alpha_{l}\vspace{-3mm}
 \label{eq:H_cq}
\eeq
and the symplectic form $i\sum_n d\bar\alpha_n\wedge d\alpha_n$. The symmetry conditions on $C_{nmkl}$ mentioned under (\ref{eq:flow_eq_linear_spectrum}) are straightforwardly understood from this expression for the Hamiltonian, and in particular they ensure that the Hamiltonian is real. Besides the Hamiltonian, the system generically admits (for any values of the interaction coefficients $C$) the following two conserved quantities:
\begin{align}
&N  =  \sum_{n=0}^{\infty}|\alpha_n|^2  \label{eq:N_cq},\\
&J =  \sum_{n=0}^{\infty}n|\alpha_n|^2  \label{eq:J_cq}.
\end{align}
Additional structures may arise for specific values of $C$, which shall be the main topic of our article. For example, extra conserved quantities have been known for the resonant systems emerging from the Gross-Pitaevskii equation \cite{GHT,BBCE,BBCE2}. An extreme case is the cubic Szeg\H o equation \cite{GG}, which corresponds to $C_{nmkl}=1$. This equation is Lax-integrable and has been thoroughly analysed \cite{GG1,GG2,GG3} with very interesting results for its dynamics.

In a good number of cases previously studied in the literature  \cite{CF,BMP,BBCE,BEL,BBCE2}, it was observed that, for some specific values of $C_{nmkl}$, (\ref{eq:flow_eq_linear_spectrum}) can be consistently truncated to the following ansatz
\begin{equation}
\alpha_n = f_n \left(b(t) + n a(t)\right) (p(t))^{n},
\label{eq:alpha_ansatz}
\end{equation} 
where $a$, $b$ and $p$ are complex-valued functions of time, and $f_n$ are time-independent numbers. Dynamics of the resonant system reduced to this three-dimensional invariant manifold can be solved explicitly. Our first objective, to be pursued in section 2, will be to specify conditions on the interaction coefficients that provide for the existence of three-dimensional invariant manifolds of the form (\ref{eq:alpha_ansatz}), and are satisfied by all previosly known cases with such three-dimensional invariant manifolds. We shall then analyze the dynamics in this class of models within the ansatz (\ref{eq:alpha_ansatz}).

In studies of resonant systems, the question of `turbulence' is important. This is understood as excitation of $\alpha_n$ with high values of $n$ starting from initial data supported by low-lying $\alpha_n$. (The terminology comes from the PDE origin of resonant systems, since $\alpha_n$ with high $n$ correspond to amplitudes of short wavelength modes of the linearlized system, thus excitation of $\alpha_n$ with high $n$ means weakly nonlinear transfer of energy to short wave length, which is literally weak turbulence). Turbulent behaviors have been observed for the cubic Szeg\H o equation. In our context, they are absent within the ansatz  (\ref{eq:alpha_ansatz}), as we explicitly show at the end of section 2.

In section 3, we develop a stronger condition on the interaction coefficients, that, first, enforces all the results of section 2, and, second, guarantees the existence of an extra complex-valued conserved quantity. This condition is likewise satisfied by all previously analyzed resonant systems with invariant manifolds  (\ref{eq:alpha_ansatz}).
However, it has only been previously known explicitly for the resonant systems derived from the Gross-Pitaevskii equation \cite{GHT,BBCE,BBCE2}. Our analysis automatically establishes this conserved quantity for all other systems in our class, including the resonant systems previously discussed in relation to the dynamics in Anti-de Sitter space \cite{CF,BEL}, where this quantity has not been known before.

In section 4 we give the general solution of the conditions on the interaction coefficients developed in section 3, which produces a very large collection of resonant systems (with arbitrary functions appearing in the definition of the interaction coefficients) which possess all of the analytic properties presented in our work.

Finally, in  section 5, we review how the previously known solvable resonant systems fit into our general framework.


\section{Three-dimensional invariant manifolds}\label{3dinv}

We are now going to formulate the condition for the resonant system (\ref{eq:flow_eq_linear_spectrum})  to have solutions of the form (\ref{eq:alpha_ansatz}). To this end, we first introduce the following definitions:
\begin{equation}\label{defS}
S_{nmkl} = \begin{cases}f_nf_mf_kf_l C_{nmkl} & n,m,k,l\ge 0, \\
\hspace{1cm}0 & \text{if any index is negative },
\end{cases}
\end{equation}
\begin{equation}
g_p^{(n,m)} \equiv \sum_{k=0}^{n+m}k^p\frac{f_kf_{n+m-k}}{f_nf_m}C_{nmk,n+m-k} = \frac{1}{f_n^2f_m^2} \sum_{k=0}^{n+m}k^pS_{nmk,n+m-k}, 
\label{eq:definition_g}
\end{equation}
\begin{equation}
F_{p}(x) \equiv \sum_{k = 0}^{\infty} k^{p} f_k^2x^k.
\label{eq:F_definition}
\end{equation}
We claim that for the closure of the ansatz (\ref{eq:alpha_ansatz}) it is sufficient to impose, with some $c_0, c_1, c_2$,
\begin{equation}
g_{0}^{(n,m)}= 1, \qquad g_{2}^{(n,m)}= c_2(n^2+m^2) + c_1 n m + c_0 (n+m)
\label{eq:summation_id_0_2}.
\end{equation}
This pattern is in fact responsible for the closure of (\ref{eq:alpha_ansatz}) in all relevant cases that emerged in the recent literature. We shall now explore the consequences of (\ref{eq:summation_id_0_2}), explain how it provides for the closure of the ansatz (\ref{eq:alpha_ansatz}), and analyze the dynamics on the three-dimensional invariant manifold parametrized by $a$, $b$, and $p$ of  (\ref{eq:alpha_ansatz}).

While for ansatz closure it is sufficient to enforce the summation identities in the form (\ref{eq:summation_id_0_2}), in fact imposing the condition on $g_0$ restricts the values of $c_0$, $c_1$ and $c_2$. This can be seen by exploring $g_0^{(n,m)}$ and $g_2^{(n,m)}$ for low-lying values of $n$ and $m$. First of all, $g^{(0,0)}_0=1$ is identical to (\ref{C0000}), which we have already assumed. Then, $g^{(0,1)}_0=1$ enforces
\beq
C_{0101}=\frac12.
\label{C0101}
\eeq
Additionally, $g_0^{(1,1)}=1$ and  $g_0^{(2,0)}=1$ enforce
\beq
2\sqrt{\gamma}\,C_{1120}+C_{1111}=1, \qquad 2\beta+\frac{1}{\sqrt{\gamma}}C_{1120}=1,
\eeq
where we have introduced
\begin{equation}
\beta = C_{2020}, \qquad \gamma = \left(\frac{f_{0}f_2}{f_1^{2}}\right)^2.
\label{defbega}
\end{equation}
Computing $g_{2}^{(n,m)}$ at low-lying $n$ and $m$ from the above expressions, we obtain
\beq
g^{(0,1)}_2=g^{(1,0)}_2=\frac12,\qquad g^{(0,2)}_2=g^{(2,0)}_2=1+2\beta,\qquad g^{(1,1)}_2=1+2\gamma(1-2\beta).
\label{g2bdry}
\eeq
From this and (\ref{eq:summation_id_0_2}),
\begin{equation}\label{g2betagamma}
g_{2}^{(n,m)} = \beta (n^2+m^2) + 2\gamma(1-2\beta) nm + \frac{1-2\beta}{2}(n+m).
\end{equation}
This is the form that we shall generally use in the computations below. Additionally, we introduce 
\beq\label{Gdef}
G\equiv\frac1{2\gamma-1},
\eeq
which will appear ubiquitously.

We now formulate \textbf{Proposition 2.1}: \textit{if} (\ref{eq:summation_id_0_2}) \textit{is satisfied and $C$ respects the symmetries listed under} (\ref{eq:flow_eq_linear_spectrum})\textit{, then $f_n$ is given by} 
\begin{equation}
f_n \equiv \begin{cases} 
\dsty\frac{1}{\sqrt{n!}} & \text{if }\gamma =\frac{1}{2},\vspace{3mm}\\
\dsty\sqrt{ \frac{\left(G\right)_n}{n!}} & \text{if }\gamma > \frac{1}{2}, 
\end{cases}
\label{eq:f_n_function}
\end{equation} 
\textit{where $(x)_n\equiv x(x+1)\cdots(x+n-1)$ denotes the rising Pochhammer symbol.} (Note that, for integer $G$, the square of the second line can be equivalently written as ${(n+1)_{G-1}}/{(G-1)!}$ which manifestly shows that it is a polynomial of degree $G-1$ in $n$.)

\vspace{0.2cm}

\noindent \textbf{Proof:} The idea behind the proof is that the summation identities (\ref{eq:summation_id_0_2}) do not manifestly respect the symmetries of $C$, and imposing the symmetries of $C$ produces extra conditions which, in particular, completely fix $f_n$. Due to the symmetries of $C$, and the corresponding symmetries of $S$ inherited through (\ref{defS}), the following identity holds for any $x$:
\beq
\sum_{n+m=k+l} \hspace{-3mm} n^2 x^{(n+m+k+l)/2} S_{nmkl}=\hspace{-3mm}\sum_{n+m=k+l} \hspace{-3mm}k^2 x^{(n+m+k+l)/2} S_{nmkl}.
\eeq
Equivalently,
\beq
\sum_{n,m=0}^\infty n^2 x^{n+m} \sum_{k=0}^{n+m} S_{nmk,n+m-k}=\sum_{n,m=0}^\infty x^{n+m} \sum_{k=0}^{n+m} k^2 S_{nmk,n+m-k}.
\eeq
Then, from (\ref{eq:definition_g}), (\ref{eq:summation_id_0_2}) and (\ref{g2betagamma}),
\beq
\sum_{n,m=0}^\infty n^2 f_n^2 f_m^2 x^{n+m} =\sum_{n,m=0}^\infty f_n^2 f_m^2 x^{n+m}\left[\beta (n^2+m^2) + 2\gamma(1-2\beta)nm + \frac{1-2\beta}{2}(n+m)\right] .
\eeq
Rewriting through (\ref{eq:F_definition}),
\beq
F_0F_2-2\gamma F_1^2-F_0F_1=0.
\eeq
We note that 
\beq\label{fpf0}
F_p=(x\del_x)^p F_0, 
\eeq
which gives the following differential equation for $F_0$:
\beq
F_0\,\del_x^2F_0-2\gamma (\del_x F_0)^2=0.
\eeq
or
\beq
\del_x(\del_x F_0/F_0)-(2\gamma-1)(\del_x F_0/F_0)^2=0,
\label{F0eq}
\eeq
which is integrated to
\beq
\frac{\del_x F_0}{F_0}=\frac{G}{h_1-x},
\eeq
and then to
\beq\label{F0ne12}
F_0=h_2\left(h_1-x\right)^{-G},
\eeq
where $h_1$ and $h_2$ are arbitrary integration constants.
If $\gamma=1/2$, then (\ref{F0eq}) is integrated to
\beq\label{F012}
F_0=h_2 \,e^{h_1x}.
\eeq
We note that if $f_n$ has been used for defining the ansatz (\ref{eq:alpha_ansatz}), then ${\tilde f}_n=d_1d_2^nf_n$ with arbitrary $d_1$ and $d_2$ is equally good, since it simply amounts to a multiplicative rescaling of $a$, $b$ and $p$. This freedom implies the possibility to arbitarily scale $F_0$ and $x$ in (\ref{F0ne12}-\ref{F012}) without essentially changing the ansatz (\ref{eq:alpha_ansatz}), and in particlar one can set $h_1$ and $h_2$ to 1, which gives
\begin{equation}
F_0(x) = \begin{cases} 
\dsty \hspace{8mm}e^x& \text{if }\gamma =\frac{1}{2},\vspace{3mm}\\
\dsty \left(1-x\right)^{-G} & \text{if }\gamma > \frac{1}{2}. 
\end{cases}
\label{eq:F0_function}
\end{equation} 
Expanding these expressions in powers of $x$ reproduces (\ref{eq:f_n_function}). For $\gamma<1/2$ (negative $G$) the Taylor coefficients of $F_0$ can be negative, and hence $f_n$ can be imaginary. Since our construction assumes real $f_n$ (and all known cases in the recent literature arising as resonant systems of the equations of mathematical physics are in this category), we shall simply discard such values of $\gamma$ and restrict ourselves to $\gamma\ge 1/2$.

\noindent\rule{5cm}{0.4pt}

Before proceeding to analyse the closure of our ansatz, we need \textbf{Proposition 2.2}: \textit{$g_0^{(n,m)}=1$ implies} 
\beq\label{eqg1}
g_1^{(n,m)}=\frac{n+m}2.
\eeq
\noindent \textbf{Proof:} The expression for $g_1^{(n,m)}$ straightforwardy follows from the symmetry of $S$ under permutation of the third and the fourth index:
\begin{align}
&g_1^{(n,m)}=\frac{1}{f_n^2f_m^2} \sum_{k=0}^{n+m}kS_{nmk,n+m-k}=\frac{1}{2f_n^2f_m^2}\left[\sum_{k=0}^{n+m}kS_{nmk,n+m-k}+\sum_{k=0}^{n+m}(n+m-k)S_{nm,n+m-k,k}\right]\nonumber\\
&\hspace{2cm}=\frac{n+m}{2f_n^2f_m^2}\sum_{k=0}^{n+m}S_{nmk,n+m-k}=\frac{n+m}2 g_0^{(n,m)}=\frac{n+m}2.
\end{align}

\noindent\rule{5cm}{0.4pt}

We now come to the main point of this section, expressed by \textbf{Proposition 2.3}: \textit{if }(\ref{eq:summation_id_0_2})\textit{ is obeyed, the ansatz }(\ref{eq:alpha_ansatz}) \textit{is respected by }(\ref{eq:flow_eq_linear_spectrum})\textit{, provided that three ordinary differential equations specified below are satisfied by $a(t)$, $b(t)$ and $p(t)$.}

\vspace{0.2cm}
\noindent\textbf{Proof:}  Substitution of  (\ref{eq:alpha_ansatz}) into (\ref{eq:flow_eq_linear_spectrum}) results in 
\beq\label{ansatzsubst}
i\left(\dot b +\dot a n +(b+an)n\frac{\dot p}{p}\right)=\sum_{j=0}^\infty f_j^2 x^j (\bar b+ \bar aj)\sum_{k=0}^{n+j} \frac{f_k f_{n+j-k}}{f_nf_j} C_{njk,n+j-k} (b+ak)(b+a(n+j-k)),
\eeq
where we have defined 
\beq\label{defx}
x = |p|^2.
\eeq 
The sum over $k$ is evaluated using  (\ref{eq:summation_id_0_2}) and (\ref{eqg1}) which follows from (\ref{eq:summation_id_0_2}), and it is a quadratic polynomial in $n$ and $m$. Summation over $j$ is then expressed in terms of $F_p$ defined by (\ref{eq:F_definition}) and given by (\ref{fpf0}) and (\ref{eq:F0_function}). The remaining expression on the right-hand side of (\ref{ansatzsubst}) is then a quadratic polynomial in $n$. The left-hand side of (\ref{ansatzsubst}) is likewise a quadratic polynomial in $n$, and equating the three coefficients of the two polynomials produces three ordinary differential equations for the three functions $a$, $b$, $p$, as claimed.

Direct evaluation of the sums in (\ref{ansatzsubst}) along the lines described above produces the following equations:
\begin{align}
i\,\dot{b} = & b^2 (\bar{b}F_0 + \bar{a}F_1) + a b (\bar{b}F_1 + \bar{a} F_2) -\frac{1-2\beta}{2} a^2\Big[\bar{b} (F_1-F_2) - \bar{a} (F_2 - F_3)\Big],\label{eqb} \\
i\,\frac{\dot{a}}a = & \frac{1+2\beta}{2}b (\bar{b}F_0 + \bar{a}F_1)  + \frac{1}{2}a \bar{b}\Big[(2\beta-1)F_0+(2-4\gamma + 8 \beta\gamma)F_1\Big]\label{eqa}\\
&\hspace{2cm} + \frac{1}{2}a \bar{a}\Big[(2\beta-1)F_1 + (2-4\gamma+8\beta\gamma)F_2\Big], \nonumber\\
i\,\frac{\dot{p}}p= & \frac{(1-2\beta)}{2}a  (\bar{b}F_0 + \bar{a}F_1),\label{eqp}
\end{align}
where $F_p$ is understood as $F_p(x)$ in all expressions.

\noindent\rule{5cm}{0.4pt}

The equations above must evidently respect the general conserved quantities of the resonant system given by (\ref{eq:H_cq}), (\ref{eq:N_cq}) and (\ref{eq:J_cq}). Restricted to our ansatz, these quantities take the form
\begin{align}
N = & |b|^2F_0 + (\bar{a}b+a\bar{b})F_1 + |a|^2F_2, \\
J = & |b|^2F_1 + (\bar{a}b+a\bar{b})F_2 + |a|^2F_3,\\
H = & N^2 + 2\gamma\left(2\beta-1\right)S^2,
\end{align}
where
\beq
S =  |a|^2 \frac{F_1}{F_0}\left(F_0 + (2\gamma-1)F_1\right).\\
\eeq
In verifying various conservation laws, one can conveniently use
\beq
\dot F_p=\frac{\dot x}{x} F_{p+1}.
\eeq

It turns out that (\ref{eqb}-\ref{eqp}) admit an extra complex conserved quantity given by
\begin{align}
&Z =  \bar{p} (|b|^2 + a \bar{b})\left(F_0 + (2\gamma-1)F_1\right) \label{defZ3d}\\
&\hspace{3cm}+ \bar{p} (|a|^2 + \bar{a}b + a\bar{b})\left(F_1 + (2\gamma-1)F_2\right) +  \bar{p}|a|^2 \left(F_2 + (2\gamma - 1)F_3\right).\nonumber
\end{align}
The conservation of $Z$ can be verified directly with (\ref{eqb}-\ref{eqp}). One can view $Z$ as descending from $\sum_n \sqrt{(n+1)(n + G)}\,\bar{\alpha}_{n+1} \alpha_n$, 
with $G$ given by (\ref{Gdef}), restricted to our ansatz (\ref{eq:alpha_ansatz}).
Note that at this point we make no claims about the conservation of this quantity by the full resonant system  (\ref{eq:flow_eq_linear_spectrum}), though this will form the main topic of the next section. 

Even though one can solve (\ref{eqb}-\ref{eqp}) explicitly using the conservation laws, this solution is not very important for us here, and we limit ourselves to \textbf{Proposition 2.4} expressing the most prominent feature of the corresponding motion: \textit{ for all solutions of} (\ref{eqb}-\ref{eqp}) \textit{, $x$ defined by} (\ref{defx})  \textit{is a periodic function of time; correspondingly the spectrum $|\alpha_n|^2$ is periodic in time.}

\vspace{0.2cm}
\noindent\textbf{Proof:} We shall essentially follow the strategy of \cite{CF}. One starts by expressing $|a|^2$, $|b|^2$ and $\Re(\bar{a}b)$ in terms of $x$ and the conserved quantities $N$, $J$ and $S$:
\begin{align}
&|a|^2  =  \frac{F_0}{F_1}\frac{1}{(F_0+(2\gamma-1)F_1)}S \label{eq:mod_a}\\
&|b|^2  = \frac{F_3 N-F_2 J}{F_1^2-F_0F_2}-\frac{F_3F_1-F_2^2}{(F_1^2-F_0F_2)^2}\left(F_1 N - F_0 J - S \frac{F_0}{F_1}\frac{F_2F_1-F_0F_3}{F_0+(2\gamma-1)F_1}\right) \label{eq:mod_b}\\
&2\Re(\bar{a}b)  =  \frac{1}{F_1^2-F_0F_2}\left(F_1 N - F_0 J - S \frac{F_0}{F_1}\frac{F_2F_1-F_0F_3}{F_0 + (2\gamma-1)F_1}\right) \label{eq:rF_ab}
\end{align}
The equation of $\dot{x}$ can also  be rewritten in terms of conserved quantities. First,
\begin{equation}
\frac{\dot{x}}x = (2\beta-1)\Im(\bar{a}b)F_0.
\end{equation}
Using $\left(\Im(\bar{a}b)\right)^2 = |a|^2|b|^2-\left(\Re(\bar{a}b)\right)^2$, one expresses the square of the right-hand side through the conserved quantities and $x$. Finally, introducing
\begin{equation}
y = \frac{x}{1-x}
\label{eq:x_to_y}
\end{equation}
one obtains
\begin{align}\label{eq:doty_2_eq}
\frac{\dot{y}^2}{(1-2\beta)^2} = &- \frac{1}{4}\left(N^2 + 8 \gamma (2\gamma-1)S^2\right) y^2\\
& + \frac{1}{2}(2\gamma-1)\Big[S(N-4\gamma S) + J (N + 2 (2\gamma-1)S)\Big] y
- \frac{1}{4}(1-2\gamma)^2\left(J - S\right)^2.\nonumber
\end{align}
The last equation is simply in the form of energy conservation for a one-dimensional harmonic oscillator, hence all of its solutions are periodic and of the form
\begin{equation}\label{ysoln}
y = B + A\sin\left(\Omega t + \theta\right), \qquad \Omega = \frac{1-2\beta}{2}\sqrt{N^2 + 8\gamma(2\gamma-1)S^2},
\end{equation}
where
\begin{align}
A = & -\frac{1}{\Omega}\sqrt{\left(\Omega B\right)^2 - \frac{(1-2\beta)^2}{4}(1-2\gamma)^2\left(J - S\right)^2},\label{oscA}\\
B = & \frac{1}{2\Omega^2}\frac{(1-2\beta)^2}{2}(2\gamma-1)\left(S(N-4\gamma S) + J (N + 2 (2\gamma-1)S)\right).\label{oscB}
\end{align}
The periodicity of $y$ is transferred to $x$ through (\ref{eq:x_to_y}) and hence to $|\alpha_n|^2$ through (\ref{eq:mod_a}-\ref{eq:rF_ab}) and
\begin{equation}
|\alpha_n|^2 = f_n^2  x^n \left(|b|^2 + 2n\Re\left(\bar{a}b\right) + n^2|a|^2\right).
\label{alphaspectrum}
\end{equation}
This is directly analogous to the periodic dynamics of the spectrum observed in specific cases in the literature \cite{CF,BBCE,BEL,BBCE2}.

\noindent\rule{5cm}{0.4pt}

For solutions given by (\ref{ysoln}), $y$ oscillates between $y_+$ and $y_-$ given by $y_{\pm} = B\pm A$. We conclude this section with \textbf{Proposition 2.5}: \textit{for a given resonant system} (\ref{eq:flow_eq_linear_spectrum}) \textit{satisfying} (\ref{eq:summation_id_0_2}), \textit{the range of motion of $y$ described by} (\ref{ysoln}), \textit{defined by $(1+y_+)/(1+y_-)$, is uniformly bounded for all solutions within the ansatz} (\ref{eq:alpha_ansatz}). Informally, this means that the energy transfer to higher modes cannot be made more and more turbulent without limit by tuning the initial conditions (as happens for the cubic Szeg\H o equation \cite{GG}).

\vspace{0.2cm}
\noindent\textbf{Proof:} Again we follow the strategy developed in \cite{CF} for a particular special case of our present setup. We write
\begin{equation}
\frac{1+y_{+}}{1+y_{-}} = \frac{(1+y_+)^2}{(1+y_-)(1+y_+)}\leq \frac{(1+y_+ + y_-)^2}{1+y_- + y_+ + y_- y_+}.
\end{equation}
The last expression has the advantage that $y_++y_-$ and $y_+y_-$ are expressible through the coefficients in (\ref{eq:doty_2_eq}), which are algebraically simpler than (\ref{oscA}-\ref{oscB}). More specifically,
\begin{align}
 \frac{(1+y_+ + y_-)^2}{1+y_- + y_+ + y_- y_+} &= \frac{\left(N + 2 (2\gamma-1)S\right)^2 \left(N + 2 (2\gamma-1) J\right)^2}{\left(N+(2\gamma-1)J + (2\gamma-1)S\right)^2 \left(N^2 + 8\gamma(2\gamma-1)S^2\right)} \nonumber\\
&\leq  \frac{4\left(N + 2 (2\gamma-1)S\right)^2}{\left(N^2 + 8\gamma(2\gamma-1)S^2\right)}\le 4\left(1+2(2\gamma-1)\frac{S}{N}\right)^2.
\end{align}
Since
\begin{equation}
N = F_0 \Big| b + \frac{F_1}{F_0} a \Big|^2 + S \geq S, 
\end{equation}
we conclude that
\begin{equation}
\frac{1+y_{+}}{1+y_{-}} \leq 4 \left(4\gamma - 1\right)^2.
\end{equation}
Hence for a fixed resonant system (fixed $\gamma$) and fixed $y_-$, the maximal value of $|p|$ in (\ref{eq:alpha_ansatz}) is bounded from above by a fixed number, irrespectively of other initial conditions.

\noindent\rule{5cm}{0.4pt}


\section{Conserved bilinears and associated symmetries}\label{bilinear}

We have previously established that imposing the summation identities (\ref{eq:summation_id_0_2}) provides for the existence of three-dimensional invariant manifolds of the cubic resonant system (\ref{eq:flow_eq_linear_spectrum}), within which the flow is analytically solvable. We have also seen that, within these three-dimensional invariant manifolds, a complex quantity given by 
\beq
Z  = \begin{cases}
\dsty\sum_{n=0}^{\infty}\sqrt{(n+1)(n + G)}\,\bar{\alpha}_{n+1} \alpha_n \qquad & \text{if } \gamma>\frac{1}{2},\vspace{2mm}\\
\dsty\sum_{n=0}^{\infty}\sqrt{n+1}\,\bar{\alpha}_{n+1} \alpha_n \qquad & \text{if } \gamma=\frac{1}{2}.
\end{cases}
\label{eq:Z_cq}
\eeq
is conserved. We shall now explore a stronger condition one can impose on the interaction coefficientes of (\ref{eq:flow_eq_linear_spectrum}) which enforces this new conservation law for all solutions, irrespectively of whether they belong to the three-dimensional invariant manifolds. We note that conservation of $Z$ has been previously known in resonant systems related to the Gross-Pitaevskii equation \cite{GHT,BBCE,BBCE2}, but not for other systems in the large class considered here. In particular, our derivations automatically supply this extra conserved quantity for a few explicit systems in the literature for which it has been previously unknown \cite{CF,BEL}.

The condition we shall impose is 
\beq\label{Ieq0}
\Iden_{nmkl}=0,
\eeq
where $\Iden$ is defined by
\begin{equation}
\Iden_{nmkl} = \begin{cases}
\left(n -1+ G\right)S_{n-1,mkl} + \left(m - 1 + G\right)S_{n,m-1,kl}\\
\hspace{3.7cm} - (k+1)S_{nm,k+1,l} - (l+1)S_{nmk,l+1} & \text{if } \gamma>\frac{1}{2},\\
S_{n-1,mkl} + S_{n,m-1,kl} - (k+1)S_{nm,k+1,l} - (l+1)S_{nmk,l+1} & \text{if } \gamma = \frac{1}{2}.
\end{cases} 
\label{eq:I_relation}
\end{equation}
Here, $S_{nmkl}$ is given by (\ref{defS}), $G$ is given by (\ref{Gdef}) and $f_n$ is given by (\ref{eq:f_n_function}).

From direct inspection, conditions (\ref{Ieq0}-\ref{eq:I_relation}) only relate the values of $S_{nmkl}$ with the same $n+m-k-l$. The sectors with different values of $n+m-k-l$ completely decouple from each other. For that reason, we can choose to only look at $n+m=k+l+1$, which contains all the interaction coefficients in our equation (\ref{eq:flow_eq_linear_spectrum}) -- and having (\ref{Ieq0}-\ref{eq:I_relation}) satisfied in this sector is sufficient to have all of our subsequent conclusions on the dynamics of (\ref{eq:flow_eq_linear_spectrum}) hold. Or we may as well impose (\ref{Ieq0}-\ref{eq:I_relation}) for all values of $n,m,k,l$, as we shall do in the next section, since the values of $S_{nmkl}$ for $n+m\ne k+l+1$ do not contribute to the dynamics of (\ref{eq:flow_eq_linear_spectrum}), and may be chosen as we wish -- in particular, may be chosen to satisfy (\ref{Ieq0}-\ref{eq:I_relation}).

 We start by proving \textbf{Proposition 3.1}: \textit{conditions} (\ref{Ieq0}-\ref{eq:I_relation}) \textit{imply} (\ref{eq:summation_id_0_2}), \textit{and hence} (\ref{g2betagamma}) \textit{and} (\ref{eqg1}). \textit{Therefore, all results of the previous section apply to any resonant system satisfying} (\ref{Ieq0}-\ref{eq:I_relation}).\vspace{1mm}

\noindent \textbf{Proof:} Assume first $\gamma>1/2$. Take $l=n+m-1-k$ in (\ref{Ieq0}) and then sum over $k$ from 0 to $n+m-1$, which gives
\begin{align}
&\sum_{k=0}^{n+m-1}\Big[\left(n -1+ G\right)S_{n-1,mk,n+m-1-k} + \left(m - 1 + G\right)S_{n,m-1,k,n+m-1-k} \\
&\hspace{4cm}- (k+1)S_{nm,k+1,n+m-1-k} - (n+m-k)S_{nmk,n+m-k}\Big]=0\nonumber.
\end{align}
If one changes $k$ to $k-1$ in the first term of the second line, the two terms in the second line can be effectively combined to yield
\begin{align}
&\sum_{k=0}^{n+m-1}\Big[\left(n -1+ G\right)S_{n-1,mk,n+m-1-k} + \left(m - 1 + G\right)S_{n,m-1,k,n+m-1-k}\Big] \\
&\hspace{8cm}-  (n+m)\sum_{k=0}^{n+m}S_{nmk,n+m-k}=0\nonumber.
\end{align}
Using (\ref{eq:definition_g}), this is written as
\beq
(n-1+G)f_{n-1}^2f_m^2 g^{(n-1,m)}_0+(m-1+G)f_{n}^2f_{m-1}^2 g^{(n,m-1)}_0-(n+m)f_{n}^2f_m^2 g^{(n,m)}_0=0.
\eeq
From (\ref{eq:f_n_function}),
\begin{equation}
\left(\frac{f_n}{f_{n+1}} \right)^2 = \frac{n+1}{n+G},
\label{eq:f_n_over_f_n+1_to_2}
\end{equation}
hence
\beq
(n+m)g^{(n,m)}_0=ng^{(n-1,m)}_0+mg^{(n,m-1)}_0.
\eeq
Since $C_{0000}=S_{0000}=1$ by our choice of the time variable, $g^{(0,0)}_0=1$. Furthermore, if either $n$ or $m$ is negative $g^{(n,m)}=0$. First, fix $m=0$, obtaining
\beq
g^{(n,0)}_0=g^{(n-1,0)}_0=\cdots=g^{(0,0)}_0=1.
\eeq
Analogously,
\beq
g^{(0,m)}_0=1.
\eeq
After that, the process can be repeated by fixing $m=1$ and proving that
\beq
g^{(n,1)}_0=1,
\eeq
and further on by induction,
\beq
g^{(n,m)}_0=1
\label{g0proved}
\eeq
for all nonnegative $n$ and $m$.

Analysis of $g^{(n,m)}_2$ proceeds in a similar fashion, and we shall be proving (\ref{g2betagamma}) directly. Multiply (\ref{Ieq0}) by $k^2$, set $l=n+m-1-k$ and then sum over $k$ from 1 to $n+m-1$, which gives
\begin{align}
&\sum_{k=1}^{n+m-1}k^2\Big[\left(n -1+ G\right)S_{n-1,mk,n+m-1-k} + \left(m - 1 + G\right)S_{n,m-1,k,n+m-1-k} \\
&\hspace{4cm}- (k+1)S_{nm,k+1,n+m-1-k} - (n+m-k)S_{nmk,,n+m-k}\Big]=0\nonumber.
\end{align}
Replace $k$ by $k-1$ in the first term of the second line and combine the two terms of the second line into one sum:
\begin{align}
&\sum_{k=0}^{n+m-1}k^2\Big[\left(n -1+ G\right)S_{n-1,mk,n+m-1-k} + \left(m - 1 + G\right)S_{n,m-1,k,n+m-1-k}\Big] \\
&\hspace{6cm}-  \sum_{k=0}^{n+m}\Big[(n+m-2)k^2+k\Big]S_{nmk,n+m-k}=0\nonumber.
\end{align}
In terms of $g_p^{(n,m)}$, this is
\begin{align}
&(n-1+G)f_{n-1}^2f_m^2 g^{(n-1,m)}_2+(m-1+G)f_{n}^2f_{m-1}^2 g^{(n,m-1)}_2\\
&\hspace{4cm}-(n+m-2)f_{n}^2f_m^2 g^{(n,m)}_2-f_{n}^2f_m^2g^{(n,m)}_1=0.\nonumber
\end{align}
Since we have already proved that $g^{(n,m)}_0=1$, from Proposition 2.2, $g^{(n,m)}_1=(n+m)/2$. Then, by (\ref{eq:f_n_over_f_n+1_to_2}),
\beq
(n+m-2)g^{(n,m)}_2=ng^{(n-1,m)}_2+mg^{(n,m-1)}_2-\frac{n+m}2.
\label{eqg2nm}
\eeq
Since we have already proved (\ref{g0proved}), (\ref{g2bdry}) holds, which (together with $g_2^{(n,m)}=0$ for negative $n$ or $m$) provides complete boundary conditions for solving (\ref{eqg2nm}).

We start with $m=0$ in (\ref{eqg2nm}). The first order finite difference equation
\beq
(n-2)g_2^{(n,0)}=n g_2^{(n-1,0)} -\frac{n}2
\eeq
has a unique solution matching $g^{(2,0)}_2$ given by (\ref{g2bdry}), which is
\beq
g_2^{(n,0)}=\beta n^2+\frac{1-2\beta}2 n.
\eeq
This manifestly agrees with (\ref{g2betagamma}).

The case $m=1$ requires separate treatment, which we give explicitly for the sake of accuracy. (The proof for higher values of $m$ will proceed inductively). One gets
\beq
(n-1)g_2^{(n,1)}=n g_2^{(n-1,1)}+\beta n^2-\beta n -\frac{1}2.
\eeq
The unique solution of this first order finite difference equation matching $g^{(1,1)}_2$ given by (\ref{g2bdry}) is
\beq
g_2^{(n,1)}=\beta n^2 +n\Bigg[2\gamma(1-2\beta)+\frac{1-2\beta}2\Bigg]+\frac12,
\eeq
which agrees with (\ref{g2betagamma}).

Now assume that (\ref{g2betagamma}) holds for $m=M-1$. By the symmetry of $g^{(n,m)}_2$ under permutation of $n$ and $m$ this also fixes 
\beq\label{gMiterinit}
g^{(M-1,M)}_2= \beta ((M-1)^2+M^2) + 2\gamma(1-2\beta) M(M-1) + \frac{1-2\beta}{2}(2M+1),
\eeq
which serves as the initial condition for solving
\begin{align}
&(n+M-2)g^{(n,M)}_2-ng^{(n-1,M)}_2\label{eqgiterM}\\
&\hspace{1cm}=M\Big[ \beta (n^2+(M-1)^2) + 2\gamma(1-2\beta) n(M-1) + \frac{1-2\beta}{2}(n+M-1)\Big]-\frac{n+M}2.\nonumber
\end{align}
The right hand side is a quadratic polynomial in $n$. The left hand side is a finite difference operator acting on $g_2$ so that for any polynomial $g_2$ it returns a polynomial of the same degree. Therefore, one can look for a particular solution in the form of quadratic polynomial by substituting a general polynomial in (\ref{eqgiterM}) and matching the coefficients. This gives
\beq\label{gMsoln}
g^{(n,M)}_2= \beta (n^2+M^2) + 2\gamma(1-2\beta) nM + \frac{1-2\beta}{2}(n+M),
\eeq
which also happens to satisfy the initial condition (\ref{gMiterinit}). This means that (\ref{gMsoln}) is the unique solution of the first order finite difference equation (\ref{eqgiterM}) satisfying (\ref{gMsoln}). Hence, we have proved that $g^{(n,m)}_2$ still respects (\ref{g2betagamma}) at $m=M$. By induction, this completes our proof of (\ref{eq:summation_id_0_2}). The special case $\gamma=1/2$ is treated similarly.

\noindent\rule{5cm}{0.4pt}

We now turn to the main target of this section, expressed by  \textbf{Proposition 3.2:} \textit{conditions} (\ref{Ieq0}-\ref{eq:I_relation}) \textit{imply conservation of} (\ref{eq:Z_cq}) \textit{by the resonant system} (\ref{eq:flow_eq_linear_spectrum}).

\vspace{0.2cm}

\noindent \textbf{Proof:} From the resonant system (\ref{eq:flow_eq_linear_spectrum}), the time derivative of $Z$ is given by
\begin{align}
\dot{Z}\sim\hspace{-2mm}&\sum_{n+m=k+l}\hspace{-3mm}\sqrt{(n+1)(n + G)}\,
C_{nmkl}\bar{\alpha}_{n+1}\bar{\alpha}_m\alpha_k\alpha_l \\
&\hspace{1cm}- \hspace{-5mm}\sum_{n+m+1=k+l}\hspace{-5mm}\sqrt{(n+1)(n + G)}\,C_{n+1,mkl}\bar{\alpha}_k\bar{\alpha}_l\alpha_m\alpha_n,\nonumber
\end{align}
where the propotionality sign implies that we are omitting overall numerical factors.

It is convenient  to re-express the above formula in terms of  $\beta_n = \alpha_n/f_n $ and $S_{nkmj} = f_nf_kf_mf_j C_{nkmj}$. Using (\ref{eq:f_n_over_f_n+1_to_2}), this gives
\beq
\dot{Z}\sim\hspace{-2mm}\sum_{n+m=k+l}\hspace{-2mm}(n +G)S_{nmkl}\bar{\beta}_{n+1}\bar{\beta}_m\beta_k\beta_l - \hspace{-5mm}\sum_{n+m+1=k+l}\hspace{-3mm}(n+1)
S_{n+1,mkl}\bar{\beta}_k\bar{\beta}_l\beta_m\beta_n.
\eeq
We swap $k$ with $n$ and $l$ with $m$ in  the second sum and redefine $n$ to $n-1$ in the first sum (while remembering that $S$ identically vanishes if any of its indices is negative). This gives
\beq
\dot{Z}\sim\hspace{-4mm}\sum_{n+m=k+l+1}\hspace{-4mm}\left[(n-1 +G)S_{n-1,mkl}-(k+1)
S_{nm,k+1,l}\right]\bar{\beta}_{n}\bar{\beta}_m\beta_k\beta_l.
\eeq
Since both the summation and $\bar{\beta}_{n}\bar{\beta}_m\beta_k\beta_l$ are symmetric under permutations $(n\leftrightarrow m)$ and $(k\leftrightarrow l)$, in order for $\dot Z$ to vanish, it is sufficient to have the symmetric part of the expression in the square brackets with respect to the said permutations vanish. This is precisely identical to (\ref{Ieq0}-\ref{eq:I_relation}). The above argument is phrased  for $\gamma>1/2$. The proof for $\gamma=1/2$ proceeds in a completely analogous fashion.

\noindent\rule{5cm}{0.4pt}

For systems that abide by the $Z$-conservations, the action of the associated symmetries must evidently be respected. The infinitesimal actions are constructed by computing the Poisson brackets of $\Re{Z}$ and $\Im{Z}$ with $H$ given by (\ref{eq:H_cq}), which produces two independent infenitesimal transformations, which can be compactly expressed in terms of $\beta_n = \alpha_n/f_n $ as
\beq
\de \beta_n \sim \begin{cases}
\dsty i\big(n\beta_{n-1}+(n+G)\beta_{n+1}\big) \qquad & \text{if } \gamma>\frac{1}{2},\vspace{2mm}\\
\dsty  i\big(n\beta_{n-1}+\beta_{n+1}\big)  \qquad & \text{if } \gamma=\frac{1}{2},
\end{cases}
\eeq
and
\beq
\de \beta_n  \sim \begin{cases}
\dsty n\beta_{n-1}-(n+G)\beta_{n+1} \qquad & \text{if } \gamma>\frac{1}{2},\vspace{2mm}\\
\dsty  n\beta_{n-1}-\beta_{n+1}  \qquad & \text{if } \gamma=\frac{1}{2}.
\end{cases}
\eeq
Within the ansatz (\ref{eq:alpha_ansatz}) these infinitesimal transformations can be straightforwardly integrated to derive the following finite forms:
\begin{align}
& p\mapsto \frac{p - i\tanh\eta}{1 +i p \tanh\eta}, \label{transZ1},\\
& a\mapsto \frac{ap}{(p\cosh\eta + i \sinh\eta)(\cosh\eta - ip \sinh\eta)^{G+1}},\\
& b \mapsto  \frac{ b(1 +i p \tanh\eta) -i  a G \tanh\eta}{(1 +i p \tanh\eta)(\cosh\eta + i p \sinh\eta)^G} ,
\end{align}
and
\begin{align}
& p\mapsto \frac{p+\tanh{\xi}}{1+p \tanh{\xi} },\\
& a\mapsto \frac{ap}{(p\cosh\xi + \sinh\xi)(\cosh{\xi}+p \sinh{\xi} )^{G+1}},\\
& b \mapsto \frac{b(1+p \tanh{\xi})-aG \tanh{\xi}}{(1+p \tanh{\xi})(\cosh{\xi}+p \sinh{\xi})^G} .
\end{align}

At $\gamma=1/2$, these transformations can be written compactly in terms of one complex parameter $\zeta$:
\begin{align}
& p\mapsto p + \zeta,\\
& b \mapsto  \left(b - a p \bar{\zeta}\right)e^{- p\bar{\zeta}- {|\zeta|^2}/{2}} ,\\
&a\mapsto  \frac{a p}{p+\zeta}e^{-p\bar{\zeta}- {|\zeta|^2}/{2}}.\label{transZlast}
\end{align}

The above transformations can be used to generate dynamical solutions within the ansatz (\ref{eq:alpha_ansatz}) from stationary solutions in a manner identical to the analysis of \cite{BBCE2}. More specifically, from (\ref{defx}), (\ref{eqp}) and (\ref{defZ3d}), one finds
\begin{equation}
\frac{\dot{x}}{2 \beta -1} = \frac{\Im(p Z) }{\dsty 1+F_1/(G F_0)} .
\end{equation}
For initial data with $Z = 0$, $x$ is constant, and hence so are $|b|^2$, $|a|^2$, $Re(a \bar{b})$ and the spectrum (\ref{alphaspectrum}). Thus one has a family of stationary solutions of the form $p = q e^{-i \omega \tau}$, $b = \beta e^{-i \lambda \tau}$ and $a = \gamma e^{-i\lambda 	\tau}$, labelled by two parameters $c$ and $q$, with real-valued
\begin{align}
& \gamma= - c (1- q^2),\\
& \beta_{\pm} = \frac{c}{2} \left(1 + (2G+1)  q^2 \pm \sqrt{1 - (6 + 4 G) q^2 + q^4}\right),\\
&\lambda_{\pm} = \frac{c^2}{4(1- q^2)^G} \Big((1- q^2) (2 +  q^2 (G (2\beta-1) - 2) )\\
&\hspace{5cm}\pm (2 +  q^2 (2 + G (1-2\beta)))\sqrt{1 - (6 + 4 G)  q^2 +  q^4}\Big),\nonumber\\
&\omega_{\pm} = c^2 \frac{(2\beta-1)}{4 (1-  q^2)^{G-1}} \left(1 +  q^2 \pm \sqrt{1 - (6 + 4 G)  q^2 +  q^4}\right).
\end{align}
In the case of $\gamma = 1/2$ there is also a stationary state for the initial data with $Z=0$:
\begin{align}
& \gamma= -c,\\
& \beta_{\pm} = \frac{c}{2}\left(1+2q^2 \pm \sqrt{1-4q^2}\right),\\
&\lambda_{\pm} = \frac{c^2}{4}e^{q^2} \left(2+(2\beta-1)q^2 \pm (2 - (2\beta -1)q^2)\sqrt{1-4q^2}\right),\\
&\omega_{\pm} = \frac{c^2}{4}e^{q^2}(2\beta-1) (1\pm \sqrt{1-4q^2}).
\end{align}
Starting with such stationary solutions, the parameter $c$ allows choosing the overall scale of $a$ and $b$, and $q$ allows choosing the absolute value of $p$. Thereafter, applying (\ref{transZ1}-\ref{transZlast}) allows adjusting the magnitude and phase of  $a/b$. Furthermore, the phase rotation symmetries $\alpha_n \mapsto e^{i\theta} \alpha_n$ and $\alpha_n \mapsto e^{i\theta n} \alpha_n$  can be used to arbitrarily adjust the common phase of $a$ and $b$, and the phase of $p$, producing generic dynamical solutions within the ansatz  (\ref{eq:alpha_ansatz}).


\section{General interaction coefficients}

Equations  (\ref{Ieq0}-\ref{eq:I_relation}) are linear finite difference equations that one can hope to analyze explicitly. In fact, in this section we shall solve them in terms of the generating function corresponding to the interaction coefficients. This will completely characterize resonant systems satisfying  (\ref{Ieq0}-\ref{eq:I_relation}). Below we present the analysis for the two relevant cases in (\ref{eq:I_relation}), $\gamma>1/2$ and $\gamma=1/2$.

\subsection{$\gamma>1/2$}

We want to find solutions to
\beq
(n-1+G)S_{n-1,mkl}+(m-1+G)S_{n,m-1,kl}-(k+1)S_{nm,k+1,l}-(l+1)S_{nmk,l+1}=0
\eeq
subject to the symmetry conditions $S_{nmkl}=S_{mnkl}=S_{klnm}$.
We start by introducing the generating function for $S_{nmkl}$,
\beq
S(y,z,v,w)=\sum_{n,m,k,l=0}^{\infty} S_{nmkl} \,y^nz^mv^kw^l,
\eeq
and write (\ref{Ieq0}-\ref{eq:I_relation}) as
\begin{align}
\sum_{n,m,k,l=0}^\infty 
&\left[y(y\del_y+G)S_{n-1,mkl}y^{n-1}z^mv^kw^l+z(z\del_z+G)S_{n,m-1,kl}y^{n}z^{m-1}v^kw^l\right.\nonumber\\
&\hspace{1cm}\left.-\del_vS_{nm,k+1,l}y^{n}z^mv^{k+1}w^l-\del_vS_{nmk,l+1}y^{n}z^mv^{k}w^{l+1}\right]=0,
\end{align}
where we must enforce $S_{nmkl}=0$ for any negative values of the indices, which amounts to $S(y,z,v,w)$ being non-singular near the origin. Hence,
\beq\label{eq_S_gen}
y(y\del_y+G)S+z(z\del_z+G)S-(\del_v+\del_w)S=0.
\eeq
The general solutions of this PDE can be constructed via the method of characteristics. Introduce $\xi$ parametrizing the following curves
\beq
y=\frac1{\xi-\xi_y},\qquad z=\frac1{\xi-\xi_z},\qquad v=\xi, \qquad w=\xi-\xi_0.
\eeq
The equation then becomes
\beq
\frac{d}{d\xi}\left((\xi-\xi_y)^G(\xi-\xi_z)^G S\right)=0,
\eeq
which is directly integrated as
\beq
(\xi-\xi_y)^G(\xi-\xi_z)^GS= A(\xi_y,\xi_z,\xi_0),
\eeq
where $A$ is an arbitrary function. Since $\xi$ is simply $v$ for our simple characteristics, it is convenient to eliminate it from the formulas in favor of $v$:
\beq\label{S_sol_A}
S=\frac{A(v-1/y,v-1/z,v-w)}{(yz)^G}.
\eeq
Because of the symmetries of $S$, (\ref{eq_S_gen}) must be obeyed as well after $(y,z)$ is permuted with $(v,w)$:
\beq\label{eq_S_gen_perm}
v(v\del_v+G)S+w(w\del_w+G)S-(\del_y+\del_z)S=0.
\eeq
By an argument identical to the one given above, we must have
\beq\label{S_sol_B}
S=\frac{B(y-1/v,y-1/w,y-z)}{(vw)^G}.
\eeq
Finally, we can explore the compatibility condition of (\ref{eq_S_gen}) and (\ref{eq_S_gen_perm}) by acting with $v(v\del_v+a)+w(w\del_w+a)-(\del_y+\del_z)$ on the former and $y(y\del_y+a)+z(z\del_z+a)-(\del_v+\del_w)$ on the latter and subtracting the results, which yields simply
\beq\label{eq_S_gen_comp}
(y\del_y+z\del_z-v\del_v-w\del_w)S=0.
\eeq
Hence,
\beq\label{S_sol_C}
S=C(yv,zv,yw,zw),
\eeq
with some function $C$ (the last of the four arguments is not independent of the first three, but it is convenient to present the dependences in this symmetric way). One can check that applying further commutation operations to (\ref{eq_S_gen}), (\ref{eq_S_gen_perm}) and (\ref{eq_S_gen_comp}) does not result in any new compatibility conditions. It therefore remains to compare the three representations (\ref{S_sol_A}), (\ref{S_sol_B}) and (\ref{S_sol_C}).

We rewrite (\ref{S_sol_A}) as
\begin{align}
S=&\frac{\tilde A(v-1/y,v-1/z,v-w)}{(yz)^G\left[(1/y-v)(1/z-v)(1/y-v+(v-w))(1/z-v+(v-w))\right]^{G/2}}\nonumber\\
&=\frac{\tilde A(v-1/y,v-1/z,v-w)}{\left[(1-vy)(1-vz)(1-wy)(1-wz)\right]^{G/2}}.\label{S_A_tilde}
\end{align}
Since the denominator only depends on $vy$, $vz$, $wy$ and $wz$, and by (\ref{S_sol_C}), the entire $S$ only depends on $vy$, $vz$, $wy$ and $wz$, the newly introduced $\tilde A$ (whose arguments are explicitly indicated above) must also be expressible as a function of  $vy$, $vz$, $wy$ and $wz$ only. By a similar argument from (\ref{S_sol_B}),
\beq
S=\frac{\tilde B(y-1/v,y-1/w,y-z)}{\left[(1-vy)(1-vz)(1-wy)(1-wz)\right]^{G/2}},
\eeq
where $\tilde B$ is a function that must be expressible both through the indicated variables and through $vy$, $vz$, $wy$ and $wz$. Furthermore, by comparison with (\ref{S_A_tilde}), it must also be expressible through $v-1/y$,$v-1/z$ and $v-w$. But there is only one combination of $y$, $z$, $v$, $w$ that can be expressed through either $(vy, vz, wy, wz)$ or $(v-1/y, v-1/z, v-w)$ or $(y-1/v,y-1/w,y-z)$, namely
\beq
\frac{(1-vy)(1-wz)}{(1-vz)(1-wy)}.
\eeq
Thus, $\tilde A$ and $\tilde B$ must be functions of this ratio. Furthermore, since the interchange of $y$ and $z$ inverts this ratio while it should not change $S$, all of the constraints can be summarized by introducing an arbitrary even function $\mathcal{F}(x)=\mathcal{F}(-x)$ and writing the general $S$ as
\beq
S(y,z,v,w)=\frac{\mathcal{F}\left(\ln\left[\frac{(1-vy)(1-wz)}{(1-vz)(1-wy)}\right]\right)}{\left[(1-vy)(1-vz)(1-wy)(1-wz)\right]^{G/2}}.\label{Sgamman12sol}
\eeq

\subsection{$\gamma=1/2$}

The solution at $\gamma=1/2$ is considerably simpler. Instead of (\ref{eq_S_gen}), one has
\beq\label{eq_S_gammahalf}
(y+z)S-(\del_v+\del_w)S=0,
\eeq
whose general solutions is
\beq
S=\mathcal{G}(y,z,v-w)e^{(y+z)(v+w)/2},
\eeq
with any $\mathcal{G}$.
By symmetry under interchange of $(y,z)$ and $(v,w)$, $\mathcal{G}$ must depend on $y$ and $z$ only in combination $y-z$. We thus write
\beq
S(y,z,v,w)=\mathcal{G}(y-z,v-w) e^{(y+z)(v+w)/2},\label{Sgamma12sol}
\eeq
where $\mathcal{G}$ is non-singular at the origin, even in both arguments and symmetric under the permutation of the two arguments.  Note that in this case the solution for $S$ contains an arbitrary function of two variables rather than one variable as in (\ref{Sgamman12sol}). The smaller number of constraints on the solution in this case can already be seen at the level of (\ref{eq_S_gammahalf}), since this equation is automatically compatible with its counterpart obtained by interchanging $(y,z)$ and $(v,w)$, not resulting in any extra compatibility conditions.

\section{Previously known cases}

A number of resonant systems with three-dimensional invariant manifolds of the form (\ref{eq:alpha_ansatz}) have appeared in the recent literature  \cite{CF,BMP,BBCE,BEL,BBCE2} and we now show how they fit in our general framework.
All of these systems can be written as (\ref{eq:flow_eq_linear_spectrum}) after rescaling time and $\alpha_n$, and all of them satisfy (\ref{Ieq0}-\ref{eq:I_relation}). 
For these systems, the three-dimensional invariant manifolds of Proposition 3.1 have been known before.
However, the conservation of $Z$ given by (\ref{eq:Z_cq}) has only been known in the context of systems emerging from the Gross-Pitaevskii equation, while our Proposition 3.2 supplies this conserved quanitity automatically for all other systems.


\subsection{Conformal Flow (CF)}

This resonant system was derived in \cite{CF} starting from the conformally invariant cubic wave equation on the Einstein cylinder $\mathbb{R}\times\mathbb{S}^3$, truncated to the rotationally symmetric sector. Further studies can be found in \cite{BHP,BHP2}. In terms of our formalism, CF is described as $(\beta,\gamma) = \left(\frac{1}{3},\ \frac{3}{4}\right)$ and
\begin{equation}
 S_{nmkl} = \text{min}(n,m,k,l)+1.
\end{equation}
As already mentioned, the existence of the three-dimensional invariant manifold was discovered in \cite{CF}, but the conservation of 
$Z$ given by (\ref{eq:Z_cq}) with $G=2$ is a novel result generated as a consequence of our treatement.
 

\subsection{Lowest Landau Level (LLL) equation}

The resonant system of the two-dimensional Gross-Pitaevskii equation with a harmonic trap was derived and its truncation to the lowest Landau level states was considered in \cite{GT}. This system describes a maximally rotating  two-dimensional Bose-Einstein condensate in a harmonic trap. In \cite{BBCE}, explicit analytic solutions of the form (\ref{eq:alpha_ansatz}) were found for this system. They describe slowly modulated precession of a single condensate vortex around the center of the harmonic trap. In \cite{GGT}, static multivortex states were found. In our present language, this system corresponds to $(\beta,\gamma) = \left(\frac{1}{4},\ \frac{1}{2}\right)$ and
\begin{equation}
S_{nmkl} =   \frac{(n+m)!}{2^{n+m}n!m!k!l!}.
\end{equation}
 

\subsection{Maximally rotating scalar field on $\mathbb{S}^3$ $(\text{mr}\mathbb{S}^{3})$}

This is a trucation of the resonant system of the conformally invariant cubic wave equation on the Einstein cylinder (already mentioned in our description of the conformal flow) to maximally rotating states. In \cite{BEL}, it was found that it admits a three-dimensional invariant manifold, but the extra conserved bilinear $Z$ was not identified at that time. This system corresponds to $(\beta,\gamma) = \left(\frac{1}{3},\ 1\right)$ and
\begin{equation}
S_{nmkl} =\frac{1}{1+n+m}.
\end{equation}
 

\subsection{Maximally rotating scalar fields in $AdS_{d+1}$ $(\text{mr}AdS_{d+1})$}

This is a generalization, discussed in \cite{BEL}, of the previous constuction on the Einstein cylinder to cubic wave equations in $(d+1)$-dimensional Anti-de Sitter spacetime for a complex scalar field of mass $m$. The mass and the number of dimensions only appear through one relevant combination $\delta = \frac{d}{2}+\sqrt{\frac{d^2}{4}+m^2}$. The maximally rotating truncation results in a one-parameter family of resonant systems admitting three-dimensional invariant manifolds. In our formalism, they correspond to  $(\beta,\gamma)$ given by
\begin{equation}
\beta(\delta) = \frac{1}{2}\frac{\delta+1}{2\delta+1}, \qquad \gamma(\delta) = \frac{1}{2}\frac{\delta+1}{\delta},
\label{eq:maxi_rot_scalar_AdS_beta_gamma}
\end{equation} 
and
\begin{equation}
S_{nmkl} = \frac{\Gamma(2\delta)}{\Gamma(\delta)^2}\frac{\Gamma(n+\delta)\Gamma(k+\delta)\Gamma(m+\delta)\Gamma(l+\delta)}{\Gamma(n+1)\Gamma(k+1)\Gamma(m+1)\Gamma(l+1)}\frac{\Gamma(n+m+1)}{\Gamma(n+m+2\delta)}.
\end{equation}
The conserved bilinear $Z$ is a new result for this system.
 

\subsection{Rotating sectors of the two-dimensional Gross-Pitaevskii equation $(\text{rGP}_{2+1})$}

In \cite{BBCE2}, the resonant system of the two-dimensional Gross-Pitaevskii equation with a harmonic potential was considered, but in contrast to \cite{BBCE}, it was truncated to modes of fixed angular momentum $\mu$, rather than the maximally rotating sector of \cite{BBCE}. This again results in a one-parameter family of resonant systems with
 \begin{equation}\label{betagammarotGP}
 \beta(\mu) = \frac{3}{8}, \qquad \gamma(\mu) = \frac{1}{2}\frac{\mu+2}{\mu+1}.
 \end{equation}
(Only integer values of $\mu$ emerge from truncations of the Gross-Pitaevskii equation, but all relevant properties are preserved for general values of $\mu$).
The interaction coefficients are given by
\begin{equation}\label{SrotGP}
S_{nmkl} = 2 \frac{\Gamma(\mu+1)\Gamma\left(\frac{1}{2}\right)}{\Gamma\left(\mu+\frac{1}{2}\right)} \int_{0}^{\infty} d\rho \, e^{-2\rho}\rho^{2\mu} L_{n}^{\mu}(\rho)L_{m}^{\mu}(\rho)L_{k}^{\mu}(\rho)L_{l}^{\mu}(\rho),
\end{equation}
where $L_{n}^{\mu}$ are the associated Laguerre polynomials.
 
Because there is no practically usable closed form expression for the integral appearing in (\ref{SrotGP}), proving (\ref{Ieq0}-\ref{eq:I_relation}) is more involved in this case than in the other cases considered above. The relations (\ref{Ieq0}-\ref{eq:I_relation})  do hold nonetheless for the interaction coefficients satisfying the resonance condition in (\ref{eq:flow_eq_linear_spectrum}), as we explicitly prove in the appendix using identities for the Laguerre polynomials.


\subsection{Excited Landau levels of the two-dimensional Gross-Pitaevskii equation $(xLL_{2+1})$}

These systems have been obtained in \cite{BBCE2} by truncating the two-dimensional Gross-Pitaevskii resonant system to excited Landau levels. They are parametrized by an integer $L\ge 1$ (while the case $L=0$ is simply the lowest Landau level mentioned above), and correspond to
\beq
\beta(L)=\frac12-\frac{L-1}{4(2L-1)},\qquad \gamma(L)=\frac12.
\eeq
The interaction coefficients are again given in terms of the Laguerre polynomials:
\beq
 S_{nmkl}= \frac{2^{2L+1}(L!)^4 }{(2L)!n!m!k!l!}\int\limits_{0}^{\infty}d\rho\, e^{-2\rho} \rho^{n+m-2L}
 L_{L}^{n-L}(\rho)L_{L}^{m-L}(\rho)L_{L}^{k-L}(\rho)L_{L}^{l-L}(\rho).\label{SxLL}
\eeq
Proving (\ref{Ieq0}-\ref{eq:I_relation}) requires a certain amount of manipulations with Laguerre polynomials, which are given explicitly in the appendix.


\subsection{General picture}

\begin{figure}[h!]
	\begin{center}
		\includegraphics[width=12cm]{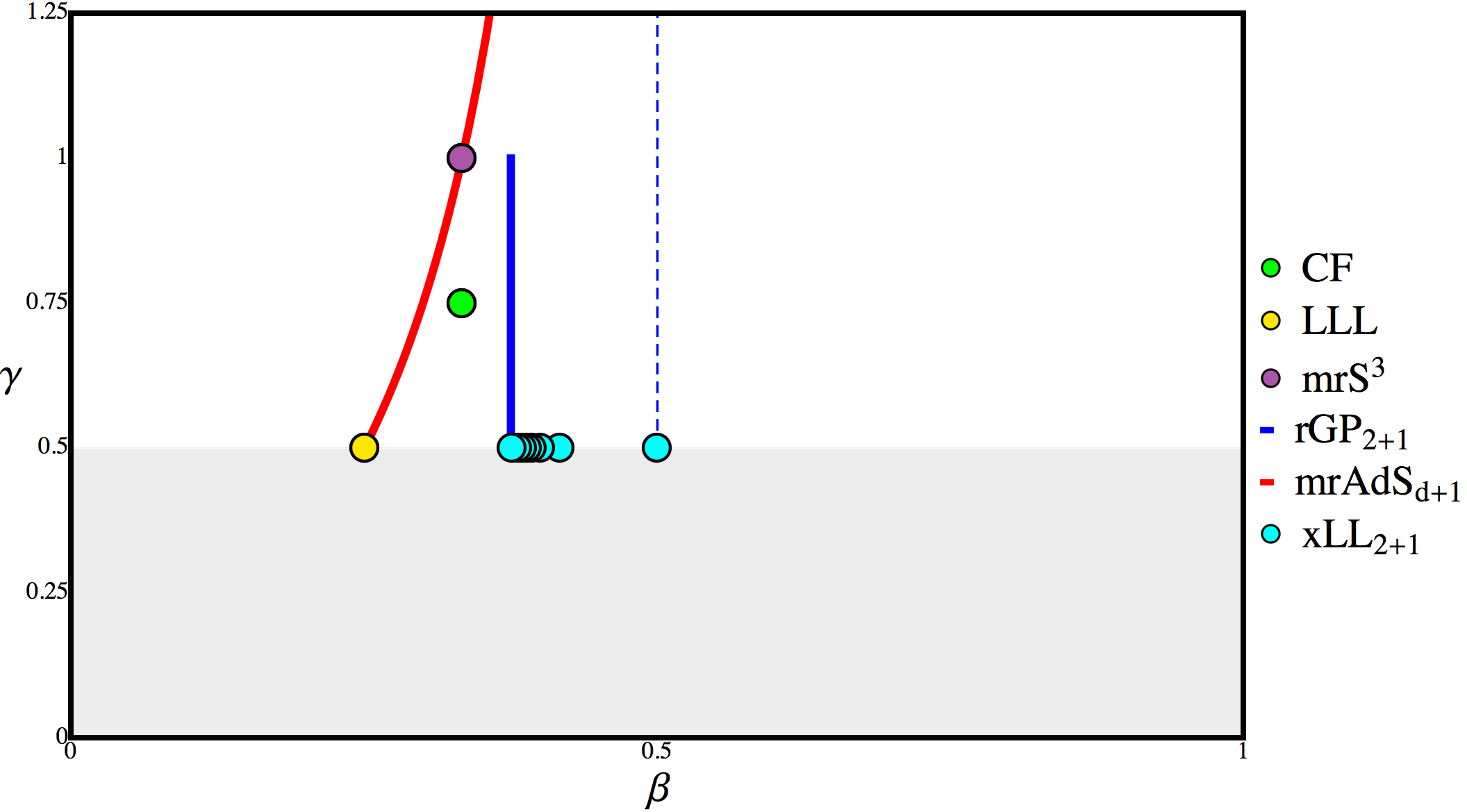}\vspace{-7mm}
	\end{center}
	\caption{\small{Currently known systems with three-dimensional invariant manifolds of the form (\ref{eq:alpha_ansatz}) arising from weakly nonlinear analysis of equations of mathematical physics, parametrised by $(\beta,\gamma)$. The vertical dashed line represents the special value $\beta=1/2$ for which the spectrum $|\alpha_n|^2$ does not evolve. We have not explored the shaded region $\gamma<1/2$, since $f_n$ of (\ref{eq:alpha_ansatz}) is assumed real in our construction, and this assumption does not work in the shaded region.}}
	\label{fig1}
\end{figure}

An important feature extracted in section \ref{3dinv} from the study of three-dimensional invariant manifolds of the form (\ref{eq:alpha_ansatz}) is that their dynamics is fully determined by $\beta$ and $\gamma$. This allows us to compactly summarize the previous literature in figure \ref{fig1}. 

We note that $\beta$ has little impact on the dynamics within the invariant manifold (\ref{eq:alpha_ansatz}), since it be scaled out from the reduced equation (\ref{eq:doty_2_eq}) by redefining time. This means that for solutions of the form (\ref{eq:alpha_ansatz}), the spectrum $|\alpha_n|^2$ only depends on $\beta$ through overall time scaling, while the phases of $\alpha_n$ may be $\beta$-sensitive. The value $\beta=1/2$ is special, as in this case the spectrum does not evolve in time at all, as is evident from (\ref{eq:doty_2_eq}). This case is represented by the first excited Landau level of the two-dimensional Gross-Pitaevskii equation \cite{BBCE2}.

We also note that $\beta$ does not appear in the conditions  (\ref{Ieq0}-\ref{eq:I_relation}), and therefore for any two resonant systems with the same value of $\gamma$, one can take a linear combination of their interaction coefficients $C_{nmkl}$ satisfying (\ref{Ieq0}-\ref{eq:I_relation}) to obtain another system of the same type.


\section{Summary}

We have devised large classes of interaction coefficients $C_{nmkl}$ in cubic resonant systems (\ref{eq:flow_eq_linear_spectrum})  for which they admit special solutions of the form (\ref{eq:alpha_ansatz}). Imposing the summation identities (\ref{eq:summation_id_0_2}) guarantees that our ansatz is consistent. The resulting reduced three-dimensional dynamical system can furthermore be analyzed, and we have established periodic exact returns of the energy spectrum to the initial configuration (Proposition 2.4) and the absence of turbulent behaviors in this sector (Proposition 2.5).

We have furthermore explored stronger conditions (\ref{Ieq0}-\ref{eq:I_relation}) on the interaction coefficients, which imply (\ref{eq:summation_id_0_2}), and provide an additional conserved quantity (\ref{eq:Z_cq}) for the resonant system (\ref{eq:flow_eq_linear_spectrum}). In practice, these considerations supply this quantity for concrete resonant systems emerging from interesting equations of mathematical physics \cite{CF,BEL}, for which it was unknown before.
The conditions (\ref{Ieq0}-\ref{eq:I_relation})  are simple enough to analyze them in full generality and obtain solutions for the interaction coefficients in terms of their generating functions, given by (\ref{Sgamman12sol}) and (\ref{Sgamma12sol}).

Our explicit characterization of a very large class of resonant systems with the above special properties opens pathways for exploring these systems beyond the ansatz  (\ref{eq:alpha_ansatz}), in particular in search of more elaborate stationary solutions that are known to exist in special cases \cite{CF,GGT}. Explicit formulas for the generating functions of the interaction coefficients (\ref{Sgamman12sol}) and (\ref{Sgamma12sol}) will enable application of complex plane methods that are likewise known to be fruitful from the past analysis of special cases. Through their connection to weakly nonlinear analysis of concrete PDEs, our resonant systems are relevant to many areas of physics: dynamics of Bose-Einstein condensates (via the Gross-Pitaevskii equation), nonlinear dynamics in Anti-de Sitter spacetime, including its relativistic gravitational dynamics, and topics in high-energy theory, to which dynamics of Anti-de Sitter spacetime connects through the program of gravitational holography. It would be interesting to explore such applications.


\section*{Acknowledgments}

We thank Ben Craps, Javier Mas and Alexandre Serantes for discussions. This research has been supported by FPA2014-52218-P from Ministerio de Economia y Competitividad, by Xunta de Galicia ED431C 2017/07, by European Regional Development Fund (FEDER), by Grant Mar\'ia de Maetzu Unit of Excellence MDM-2016-0692, by Polish National Science Centre grant number 2017/26/A/ST2/00530 and by CUniverse research promotion project by Chulalongkorn University (grant CUAASC). A.B. thanks the Spanish program ``ayudas para contratos predoctorales para la formaci\'on de doctores 2015'' and its mobility program for his stay at Jagiellonian University, where part of this project was developed.

\appendix

\section*{Appendix: Identities for the interaction coefficients of the Gross-Pitaevskii resonant system}

We start by analyzing the fixed angular momentum sector of the Gross-Pitaevskii resonant system. We would like to show that the  interaction coefficients (\ref{SrotGP}) satisfy (\ref{Ieq0}-\ref{eq:I_relation}) provided that $n+m=k+l+1$, which are the only values of the indices in (\ref{Ieq0}-\ref{eq:I_relation}) for which the conditions constrain the interaction coefficients actually appearing in the resonant system (\ref{eq:flow_eq_linear_spectrum}).

From (\ref{betagammarotGP}), $G=\mu+1$ and therefore 
\begin{align}
&\Iden_{nmkl}\sim\int_0^\infty d\rho\, e^{-2\rho} \rho^{2\mu} \big[(n+\mu)L_{n-1}^{\mu}L_{m}^{\mu}L_{k}^{\mu}L_{l}^{\mu}+(m+\mu)L_{n}^{\mu}L_{m-1}^{\mu}L_{k}^{\mu}L_{l}^{\mu}\\
&\hspace{4cm}-(k+1)L_{n}^{\mu}L_{m}^{\mu}L_{k+1}^{\mu}L_{l}^{\mu}-(l+1)L_{n}^{\mu}L_{m}^{\mu}L_{k}^{\mu}L_{l+1}^{\mu}\big].\nonumber
\end{align}
We now use
\beq
nL_n^\mu=(n+\mu)L_{n-1}^\mu-\rho L_{n-1}^{\mu+1}
\label{lagiter1}
\eeq
in all four terms to rewrite the above as
\begin{align}
&\Iden_{nmkl}\sim\int_0^\infty d\rho\, e^{-2\rho} \rho^{2\mu} \big[(n+m-k-l-2\mu-2)L_{n}^{\mu}L_{m}^{\mu}L_{k}^{\mu}L_{l}^{\mu}\\
&\hspace{3cm}+\rho(L_{n-1}^{\mu+1}L_{m}^{\mu}L_{k}^{\mu}L_{l}^{\mu}+L_{n}^{\mu}L_{m-1}^{\mu+1}L_{k}^{\mu}L_{l}^{\mu}+L_{n}^{\mu}L_{m}^{\mu}L_{k}^{\mu+1}L_{l}^{\mu}+L_{n}^{\mu}L_{m}^{\mu}L_{k}^{\mu}L_{l}^{\mu+1})\big].\nonumber
\end{align}
Thereafter, we use $n+m-k-l=1$ in the first line and
\beq
L_n^\mu=L_n^{\mu+1}-L_{n-1}^{\mu+1}
\label{lagiter2}
\eeq
in the last two terms of the second line to obtain
\begin{align}
&\Iden_{nmkl}\sim\int_0^\infty d\rho\, e^{-2\rho} \rho^{2\mu} \big[(2\rho-2\mu-1)L_{n}^{\mu}L_{m}^{\mu}L_{k}^{\mu}L_{l}^{\mu}\\
&\hspace{3cm}+\rho(L_{n-1}^{\mu+1}L_{m}^{\mu}L_{k}^{\mu}L_{l}^{\mu}+L_{n}^{\mu}L_{m-1}^{\mu+1}L_{k}^{\mu}L_{l}^{\mu}+L_{n}^{\mu}L_{m}^{\mu}L_{k-1}^{\mu+1}L_{l}^{\mu}+L_{n}^{\mu}L_{m}^{\mu}L_{k}^{\mu}L_{l-1}^{\mu+1})\big].\nonumber
\end{align}
Finally, we use
\beq
\del_\rho L_n^\mu=-L_{n-1}^{\mu+1}
\label{lagiter3}
\eeq
to obtain
\beq
\Iden_{nmkl}\sim\int_0^\infty d\rho\, e^{-2\rho} \rho^{2\mu} (2\rho-2\mu-1-\rho\del_\rho)L_{n}^{\mu}L_{m}^{\mu}L_{k}^{\mu}L_{l}^{\mu},
\eeq
which is the same as
\beq
\Iden_{nmkl}\sim\int_0^\infty d\rho\,\del_\rho\Big[ e^{-2\rho} \rho^{2\mu+1} L_{n}^{\mu}L_{m}^{\mu}L_{k}^{\mu}L_{l}^{\mu}\Big]=  e^{-2\rho} \rho^{2\mu+1} L_{n}^{\mu}(\rho)L_{m}^{\mu}(\rho)L_{k}^{\mu}(\rho)L_{l}^{\mu}(\rho)\Big|_{\rho=0}^{\rho=\infty},
\eeq
which evidently equals 0.

For excited Landau level truncations of the Gross-Pitaevskii resonant system, we have to prove that (\ref{SxLL}) satisfy  (\ref{Ieq0}-\ref{eq:I_relation}) with $\gamma=1/2$, provided that $n+m=k+l+1$. We start by writing
\begin{align}
&n!m!k!l!\Iden_{nmkl}\sim\int_0^\infty d\rho \,e^{-2\rho} \rho^{n+m-1-2L} 
\big[nL_L^{n-1-L}L_L^{m-L}L_L^{k-L}L_L^{l-L}+mL_L^{n-L}L_L^{m-1-L}L_L^{k-L}L_L^{l-L}\nonumber\\
&\hspace{4cm}-\rho L_L^{n-L}L_L^{m-L}L_L^{k+1-L}L_L^{l-L}-\rho L_L^{n-L}L_L^{m-L}L_L^{k-L}L_L^{l+1-L}\big].
\end{align}
Applying (\ref{lagiter2}) followed by (\ref{lagiter1}) in the first line, and (\ref{lagiter2}) in the second line, we get
\begin{align}
&\Iden_{nmkl}\sim\int_0^\infty d\rho \,e^{-2\rho} \rho^{n+m-1-2L} 
\big[(n+m-2L-2\rho)L_L^{n-L}L_L^{m-L}L_L^{k-L}L_L^{l-L}\nonumber\\
&\hspace{3cm}-\rho (L_{L-1}^{n+1-L}L_L^{m-L}L_L^{k-L}L_L^{l-L}+L_L^{n-L}L_{L-1}^{m+1-L}L_L^{k-L}L_L^{l-L}\\
&\rule{0mm}{6.5mm}\hspace{3.5cm}+L_{L}^{n-L}L_L^{m-L}L_{L-1}^{k+1-L}L_L^{l-L}+L_{L}^{n-L}L_L^{m-L}L_L^{k-L}L_{L-1}^{l+1-L})\big].\nonumber
\end{align}
Due to (\ref{lagiter3}), this is the same as
\begin{align}
&\Iden_{nmkl}\sim\int_0^\infty d\rho \,\del_\rho\big[e^{-2\rho} \rho^{n+m-2L} 
L_L^{n-L}L_L^{m-L}L_L^{k-L}L_L^{l-L}\big]\\
&\hspace{5cm}=e^{-2\rho} \rho^{n+m-2L} 
L_L^{n-L}L_L^{m-L}L_L^{k-L}L_L^{l-L}\Big|_{\rho=0}^{\rho=\infty},\nonumber
\end{align}
which evidently equals 0. (Note that $n+m=k+l+1$, and the polynomials $L^{n-L}_L(\rho)$ do not include any powers of $\rho$ below $\rho^{L-n}$, which ensures that the contribution at $\rho=0$ vanishes.)


\end{document}